\documentclass[twocolumn,showpacs,preprintnumbers,amsmath,amssymb]{revtex4}

\usepackage{graphicx}
\usepackage{dcolumn}
\usepackage{bm}

\begin{document}

\preprint{}

\title{Enhancing or suppressing spin Hall effect of light in layered nanostructures}
\author{Hailu Luo}
\author{Xiaohui Ling}
\author{Xinxing Zhou}
\author{Weixing Shu}
\author{Shuangchun Wen}\email{scwen@hnu.cn}
\author{Dianyuan Fan}
\affiliation{Key Laboratory for Micro/Nano Opto-Electronic Devices
of Ministry of Education, College of Information Science and
Engineering, Hunan University, Changsha 410082, People's Republic of
China}
\date{\today}

\begin{abstract}
The spin Hall effect (SHE) of light in layered nanostructures is
investigated theoretically in this paper. A general propagation
model describing the spin-dependent transverse splitting of wave
packet in the SHE of light is established from the viewpoint of
classical electrodynamics. We show that the transverse displacement
of wave-packet centroid can be tuned to either a negative or a
positive value, or even zero, by just adjusting the structure
parameters, suggesting that the SHE of light in layered
nanostructures can be enhanced or suppressed in a desired way. The
inherent physics behind this interesting phenomenon is found to be
attributed to the optical Fabry-Perot resonance. We believe that
these findings will open the possibility for developing new
nano-photonic devices.
\end{abstract}


\maketitle

\section{Introduction}\label{SecI}
Spin Hall effect (SHE) is a transport phenomenon, in which an
applied field on the spin particles leads to a spin-dependent
displacement perpendicular to the electric field
direction~\cite{Murakami2003,Sinova2004,Wunderlich2005}. The SHE of
light can be regarded as a direct optical analogy of SHE in
electronic system where the spin electrons and electric potential
are replaced by spin photons and refractive index gradient,
respectively~\cite{Onoda2004,Bliokh2006,Hosten2008}. The SHE of
light is sometimes referred to as the Fedorov-Imbert effect, which
was predicted theoretically by Fedorov~\cite{Fedorov1965}, and
experimentally confirmed by Imbert~\cite{Imbert1972}. The
spin-dependent transverse shift in the SHE of light is generally
believed as a result of an effective spin-orbital interaction, which
describes the mutual influence of the spin (polarization) and
trajectory of the light beam~\cite{Bliokh2007}.

Recently, the SHE of light has been extensively investigated in
different physical systems. In a static gravitational field, the
photon Hamiltonian shows a new kind of helicity-torsion coupling,
resulting in a novel birefringence phenomenon: photons with distinct
helicity follow different geodesics~\cite{Gosselin2007}. In optical
systems, the SHE of light is observed directly in the glass cylinder
and its fundamental origin is related to the dynamical action of the
topological Berry-phase monopole in the evolution of
light~\cite{Bliokh2008}. The SHE of light can also be observed in
scattering from dielectric spheres~\cite{Haefner2009}. In
particular, a giant SHE of light can be produced by subwavelength
displacements of a nanoparticle~\cite{Herrera2010}. Even in free
space, the SHE of light can be observed on the direction tilted with
respect to beam propagation axis~\cite{Aiello2009}. In plasmonic
systems, a spin-dependent splitting of the focal spot of a plasmonic
focusing lens was demonstrated and explained in terms of a geometric
phase~\cite{Gorodetski2008}. In semiconductor physics, the SHE of
light has been observed in silicon via free-carrier absorption. The
interesting result suggests that the SHE of light has the potential
of probing spatial distributions of electron spin
states~\cite{Menard2010}.

The SHE may offer an effective way to manipulate the spin particles,
and open a promising way to some potential applications, such as in
dense data storage, ultra-fast information processing, and even
quantum computing~\cite{Wolf2001,Chappert2007,Awschalom2007}. The
generation, manipulation, and detection of spin-polarized electrons
in semiconductors and nanostructures define the main challenges of
spin-based electronics. Similar challenges also exist in spin-based
photonics. The SHE of light may open new opportunities for
manipulating photon spin and developing new generation of
all-optical devices as counterpart of recently presented spintronics
devices. In this paper, we will study the SHE of light in layered
nanostructures in which the refractive indices of their constituent
materials vary between high-index regions and low-index regions.
Such an environment presents to photons as an analogy of
semiconductor presenting potential to
electrons~\cite{Joannopoulos1995}, thus presenting some imaginable
interesting properties of the SHE of light.

The paper is organized as follows. First, we want to establish a
three-dimensional propagation model to describe the SHE of light in
layered nanostructure. The Fresnel coefficients are no longer real
in the layered nanostructures, so it is necessary for us to obtain a
more general expression.  Next, we attempt to reveal what roles the
Fresnel reflection and transmission coefficients play in the SHE of
light. We find that the Fresnel coefficients present sine-like
oscillations and the spin-dependent splitting of wave-packet
centroid significantly depends on their ratio. Finally, we want to
explore the secret underlying this interesting phenomenon. The
result shows that the SHE of light can be readily modulated, i.e.,
enhanced or suppressed, via tuning the optical resonance in layered
nanostructures.

\section{Three-dimensional beam propagation model}\label{SecII}
\begin{figure}
\includegraphics[width=8cm]{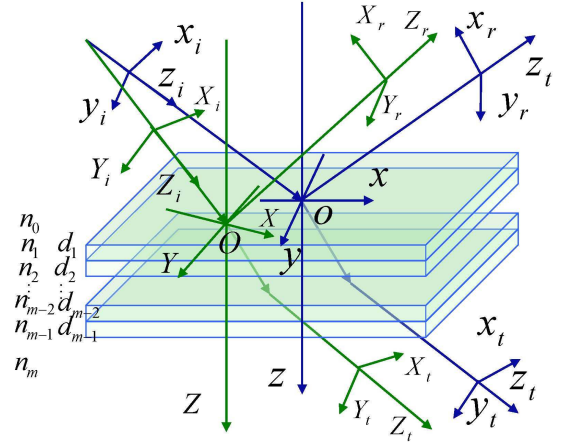}
\caption{\label{Fig1} (color online) Schematic illustrating the
reflection and refraction of central and local wave vectors at
$(m+1)$-layered nanostructure. $x_ay_az_a$ ($a=i,r,t$) are reference
frames for central wave vector of incident, reflection, and
transmission, respectively. $oxyz$ and $OXYZ$ are the interface
reference frames for central and local wave vectors, respectively.}
\end{figure}

Figure~\ref{Fig1} illustrates the beam reflection and refraction in
the layered nanostructure. The $z$ axis of the laboratory Cartesian
frame ($x,y,z$) is normal to the interfaces of the layered
structure. We use the coordinate frames ($x_a,y_a,z_a$) for central
wave vector, where $a=i,r,t$ denotes incident, reflection, and
transmission, respectively. We apply the angular spectrum method to
derive an expression for a three-dimensional beam propagation model.
Hence, we use local Cartesian frames ($X_a,Y_a,Z_a$) to describe an
arbitrary angular spectrum. The electric field of the $a$th beam can
be solved by employing the Fourier transformations. The complex
amplitude for the $a$th beam can be conveniently expressed
as~\cite{Goodman1996}
\begin{eqnarray}
\mathbf{E}_a(x_a,y_a,z_a )&=&\int d k_{ax}dk_{ay}
\tilde{\mathbf{E}_a}(k_{ax},k_{ay})\nonumber\\
&&\times\exp [i(k_{ax}x_a+k_{ay}y_a+ k_{az} z_a)],\label{noapr}
\end{eqnarray}
where $k_{az}=\sqrt{k_a^2-(k_{ax}^2+k_{ay}^2)}$ and
$\tilde{\mathbf{E}_a}(k_{ax},k_{ay})$ is the angular spectrum. The
approximate paraxial expression for the field in Eq.~(\ref{noapr})
can be obtained by the expansion of the square root of $k_{az}$ to
the first order~\cite{Lax1975}, which yields
\begin{eqnarray}
\mathbf{E}_a&=&\exp(i k_a z_a) \int dk_{ax}dk_{ay}
\tilde{\mathbf{E}_a}(k_{ax},k_{ay})\nonumber\\
&&\times\exp \left[i\left(k_{ax}x_a+k_{ay}y_a-\frac{k_{ax}^2+k_{a
y}^2}{2 k_a}z_a\right)\right]\label{apr}.
\end{eqnarray}
In general, an arbitrary linear polarization can be decomposed into
horizontal and vertical components. In the spin basis set, the
angular spectrum can be written as:
\begin{equation}
\tilde{E}_i^H=\frac{1}{\sqrt{2}}(\tilde{\mathbf{E}}_{i+}+\tilde{\mathbf{E}}_{i-})\label{SBH},
\end{equation}
\begin{equation}
\tilde{E}_i^V=\frac{1}{\sqrt{2}}i(\tilde{\mathbf{E}}_{i-}-\tilde{\mathbf{E}}_{i+})\label{SBV}.
\end{equation}
Here, $H$ and $V$ represent horizontal and vertical polarizations,
respectively. The positive and negative signs denote the left and
right circularly polarized (spin) components, respectively~\cite{Beth1936}. The monochromatic
Gaussian beam can be formulated as a localized wave packet whose
spectrum is arbitrarily narrow, and can be written as
\begin{equation}
\tilde{\mathbf{E}}_{i\pm}=\frac{1}{\sqrt{2}}(\mathbf{e}_{ix} \pm
i\mathbf{e}_{iy})\frac{w_0}{\sqrt{2\pi}}\exp\left[-\frac{w_0^2(k_{ix}^2+k_{iy}^2)}{4}\right]\label{asi},
\end{equation}
where $w_0$ is the beam waist. After the angular spectrum is known,
we can obtain the field characteristics for the $a$th beam.

To accurately describe the SHE of light in layered nanostructure, it
is need to determine the reflection and transmission of arbitrary
wave-vector components, which can be solved by $2\times2$
transmission matrix~\cite{Yariv2007}:
\begin{equation}
\mathbb{M}=T_{01}P_{1}T_{12}P_{2}...P_{m-2}T_{m-2,m-1}P_{m-1}T_{m-1,m}\label{asi},
\end{equation}
where
\begin{eqnarray}
T_{m-1,m}=\frac{1}{t_{m-1,m}}\left[
\begin{array}{cc}
1 &r_{m-1,m}\\
r_{m-1,m} & 1
\end{array}
\right],\label{matrixr}
\end{eqnarray}
is the transformation matrix from $(m-1)$-th to $m$-th layer, and
\begin{eqnarray}
P_{m}=\left[
\begin{array}{cc}
\exp(ik_{mz} d_m) &0\\
0 & \exp(-ik_{mz} d_m)
\end{array}
\right],\label{matrixr}
\end{eqnarray}
is the transmission matrix for $m$-th layer. Here, $d_m$ is the
thickness of $m$-th layer, $r_{m-1,m}$ and $t_{m-1,m}$ are
reflection and transmission coefficients from $(m-1)$-th to $m$-th
layer, respectively. For an arbitrary wave-vector component, the
Fresnel coefficients of the layered nanostructures can be written as
\begin{equation}
t_{p,s}=\frac{1}{\mathbb{M}_{11}},~~~~~r_{p,s}=\frac{\mathbb{M}_{21}}{\mathbb{M}_{11}}\label{asi},
\end{equation}
where $p$ and $s$ denote parallel and perpendicular polarizations,
respectively. By making use of Taylor series expansion, the Fresnel
coefficients can be expanded as a polynomial of $k_{ix}$. We obtain
a sufficiently good approximation when the Taylor series are
confined to the zero order.

From the boundary condition, we obtain $k_{rx}=-k_{ix} $ and
$k_{ry}= k_{iy}$. After a series of calculations of the reflected
angular spectrum given in the Appendix~\ref{AppA}, Eq.~(\ref{apr})
together with Eqs.~(\ref{asi}) and (\ref{matrixrII}) provides the
paraxial expression of the reflected field:
\begin{eqnarray}
\mathbf{E}_{r\pm}^H&=&\frac{r_p(\mathbf{e}_{rx}\pm
i\mathbf{e}_{ry})}{\sqrt{\pi
}w_0}\frac{z_R}{z_R+iz_r}\exp(ik_rz_r)\nonumber\\
&&\times\exp\left[-\frac{k_0}{2}\frac{x_{r}^2+(y_{r}\pm\delta_r^H)^2}{z_R+iz_r}\right]\label{HPR},
\end{eqnarray}
\begin{eqnarray}
\mathbf{E}_{r\pm}^V&=&\frac{\mp i r_s(\mathbf{e}_{rx} \pm
i\mathbf{e}_{ry})}{\sqrt{\pi
}w_0}\frac{z_R}{z_R+iz_r}\exp(ik_rz_r)\nonumber\\
&&\times\exp\left[-\frac{k_0}{2}\frac{x_{r}^2+(y_{r}\pm\delta_r^V)^2}{z_R+iz_r}\right]\label{VPR},
\end{eqnarray}
where $z_R=k_0w_0^2/2$ is the Rayleigh lengths,
$\delta_r^H=(1+r_s/r_p)\cot\theta_i/k_0$ and
$\delta_r^V=(1+r_p/r_s)\cot\theta_i/k_0$.

We next consider the transmitted field. From the Snell's law under
the paraxial approximation, we obtain $k_{tx}=k_{ix}/\eta $ and
$k_{ty}= k_{iy}$. Substituting Eqs.~(\ref{asi}) and
(\ref{matrixtII}) into Eq.~(\ref{apr}), we obtain the transmitted
field:
\begin{eqnarray}
\mathbf{E}_{t\pm}^H&=&\frac{t_p(\mathbf{e}_{tx} \pm
i\mathbf{e}_{ty})}{\sqrt{\pi} w_0}\frac{ z_{Ry} \exp
(ik_tz_t)}{\sqrt{(z_{Rx}+iz_t)(z_{Ry}+iz_t)}}
\nonumber\\&&\times\exp\left[-\frac{n_m k_0}{2}\left(\frac{x_{t}^2}{
z_{Rx}+iz_t}+\frac{(y_{t}\mp\delta_t^H)^2}{z_{Ry}+iz_t}\right)\right]\label{HPT},
\end{eqnarray}
\begin{eqnarray}
\mathbf{E}_{t\pm}^V&=&\frac{\mp i t_s(\mathbf{e}_{tx} \pm
i\mathbf{e}_{ty})}{\sqrt{\pi} w_0}\frac{ z_{Ry} \exp
(ik_tz_t)}{\sqrt{(z_{Rx}+iz_t)(z_{Ry}+iz_t)}}
\nonumber\\&&\times\exp\left[-\frac{n_m k_0}{2}\left(\frac{x_{t}^2}{
z_{Rx}+iz_t}+\frac{(y_{t}\mp\delta_t^V)^2}{z_{Ry}+iz_t}\right)\right]\label{VPT}.
\end{eqnarray}
Here, $\delta_t^H=(\eta-t_s/t_p)\cot\theta_i/k_0$ and
$\delta_t^V=(\eta- t_p/t_s)\cot\theta_i/k_0$. The interesting point
we want to stress is that there are two different Rayleigh lengths,
$z_{Rx}=n_m\eta^2k_0w_0^2/2$ and $z_{Ry}=n_m k_0w_0^2/2$,
characterizing the spreading of the beam in the direction of $x$ and
$y$ axes, respectively~\cite{Luo2009}. Note that the Fresnel
coefficients are no longer real in the layered nanostructure. Hence,
we should extend the previous expression of transverse
displacement~\cite{Hosten2008} to a more general situation.

\section{Spin Hall effect of light}\label{SecII}
It is well known that the SHE of light manifests itself as
polarization-dependent transverse splitting. To reveal the SHE of
light, we now determine the transverse displacements of field
centroid. The time-averaged linear momentum density associated with
the electromagnetic field can be shown to be~\cite{Jackson1999}
\begin{equation}
\mathbf{p}_{a}(\mathbf{r})=
\frac{1}{2c^2}\mathrm{Re}[\mathbf{E}_{a}(\mathbf{r})
\times\mathbf{H}_{a}^\ast(\mathbf{r})]\label{LMD},
\end{equation}
where the magnetic field can be obtained by
$\mathbf{H}_{a}=-ik_{a}^{-1} \nabla\times\mathbf{E}_{a}$. The
intensity distribution of wave packet is closely linked to the
longitudinal momentum currents
$I(x_a,y_a,z_a)\propto\mathbf{p}_a\cdot \mathbf{e}_{az}$.

At any given plane $z_a=\text{const.}$, the transverse displacement
of wave-packet centroid compared to the geometrical-optics
prediction is given by
\begin{equation}
\Delta y_{a}= \frac{\int \int y_a I(x_a,y_a,z_a) \text{d}x_a
\text{d}y_a}{\int \int I(x_a,y_a,z_a) \text{d}x_a
\text{d}y_a}.\label{centroid}
\end{equation}
Note that the transverse displacement can be divided into
$z_a$-dependent and $z_a$-independent terms. We here concentrate our
attention on the $z_a$-independent transverse displacements.

We first consider the spin-dependent transverse displacement of the
reflected field. After substituting the reflected field
Eqs.~(\ref{HPR}) and (\ref{VPR}) into Eq.~(\ref{centroid}), we
obtain the transverse spatial displacements as
\begin{equation}
\Delta y_{r\pm}^H =\mp\frac{\lambda}{2\pi}
[1+|r_s|/|r_p|\cos(\varphi_s-\varphi_p)]\cot \theta_i\label{TSRH},
\end{equation}
\begin{equation}
\Delta y_{r\pm}^V
=\mp\frac{\lambda}{2\pi}[1+|r_p|/|r_s|\cos(\varphi_p-\varphi_s)]\cot
\theta_i\label{TSRV},
\end{equation}
where $r_{p,s}=|r_{p,s}|\exp(i\varphi_{p,s})$.  Note that these
expressions are slightly different from the previous
work~\cite{Hosten2008,Qin2009}, since the Fresnel reflection
coefficients are no longer real in our model.

\begin{figure}
\includegraphics[width=8cm]{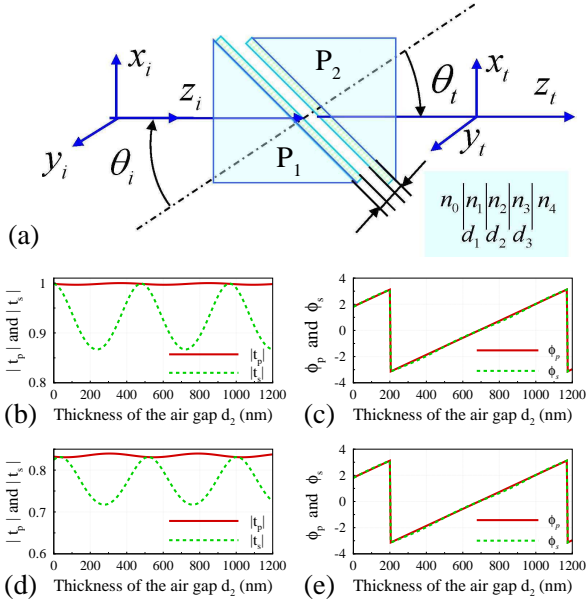}
\caption{\label{Fig2} (Color online) The Fresnel transmission
coefficients in a five-layered nanostructure. (a) Schematic
illustrating the layered nanostructure: The two prisms (P$_1$ and
P$_2$) both filmed by a low refractive index layer with
$n_1=n_3=1.377$ ($\mathrm{MgF_2}$ at 633 nm), $d_1=d_3=80 nm$, and
the thickness of the air gap ($d_2$) can be modulated. The incident
angle is chosen as $\theta_i=\pi/6$. The Fresnel transmission
coefficients $t_p$ and $t_s$ versus the thickness of the air gap
$d_2$ for the the symmetric and asymmetric system: (b)
$n_0=n_4=1.515$ (BK7 at 633 nm) and (d) $n_0=1.515$ (BK7 at 633 nm)
and $n_4=1.996$ (S-LAH79 at 633 nm). (c) and (e) are the phase of
the Fresnel transmission coefficients for the two systems,
respectively.}
\end{figure}

We next consider the spin-dependent transverse displacements of the
transmitted field. After substituting the transmitted field
Eqs.~(\ref{HPT}) and (\ref{VPT}) into Eq.~(\ref{centroid}), we have
\begin{equation}
\Delta y_{t\pm}^H =\pm\frac{\lambda}{2\pi}
[\eta-|t_s|/|t_p|\cos(\phi_s-\phi_p)]\cot \theta_i\label{TSRH},
\end{equation}
\begin{equation}
\Delta y_{t\pm}^V =\pm\frac{\lambda}{2\pi}
[\eta-|t_p|/|t_s|\cos(\phi_p-\phi_s)]\cot \theta_i\label{TSRV},
\end{equation}
where $t_{p,s}=|t_{p,s}|\exp(i \phi_{p,s})$. For an arbitrary
linearly polarized incident beam, the calculation of the transverse
displacements for the reflected and transmitted field is given in
the Appendix~\ref{AppB}.

For left and right circularly polarized components, the eigenvalues
of the transverse displacement are the same in magnitude but
opposite in directions. Under the limit of $m=1$ (air-glass
interface), the above expression coincides well with the early
results~\cite{Bliokh2007}. Our scheme shows that the SHE of light
can be explained from the viewpoint of classic electrodynamics. For
incidence angles greater than the critical angle of total internal
reflection, most of photons are reflected, and part of them tunnel
through the layered structure~\cite{Luo2010}. Hereafter, we only
concentrate our attention on the SHE of light in the transmission
case.

As an example, a five-layered nanostructure composed of two prisms
(P$_1$ and P$_2$), two films ($\mathrm{MgF_2}$), and an air gap
[Fig.~\ref{Fig2}(a)] is chosen to illustrate the SHE of light. We
consider two kind of systems: (i) A symmetric system with two BK7
prisms ($n_0=n_4=1.515$ at 633 nm) filmed by $\mathrm{MgF_2}$
($n_1=n_3=1.377$ at 633 nm) and the middle layer is air gap
($n_2=1$); (ii) Replacing P$_2$ by an S-LAH79 prism ($n_4=1.996$ at
633 nm) to form an asymmetric system. The $\mathrm{MgF_2}$ layers
with 80 nm thickness were prepared on P$_1$ and P$_2$. For a given
incident angle $\theta_i=\pi/6$, $t_s$ in the two systems both
behave sine-like oscillations versus to the thickness of air gap
($d_2$) due to the optical Febry-Perot resonance with multi-resonant
peaks for different $d_2$, while $t_p$ is nearly unchanged and
insensitive to $d_2$ [Fig.~\ref{Fig2}(b) and~\ref{Fig2}(d)]. It is
obvious that $|t_s|/|t_p|\cos(\phi_s-\phi_p)$ and
$|t_p|/|t_s|\cos(\phi_p-\phi_s)$ in Eqs.~(\ref{TSRH})
and~(\ref{TSRV}) determine the magnitude of the transverse
displacements ($\Delta y_{t\pm}$) of wave-packet centroid since
other quantities in the two equations are constant. Actually, the
$\phi_p$ is nearly equal to $\phi_s$ versus $d_2$ for both the
symmetric [Fig.~\ref{Fig2}(c)] and asymmetric system
[Fig.~\ref{Fig2}(e)], which means $\cos(\phi_p-\phi_s)\approx 1$ and
the magnitude of $\Delta y_{t\pm}$ only depends on $|t_s|/|t_p|$ or
$|t_p|/|t_s|$.

\begin{figure}
\includegraphics[width=8cm]{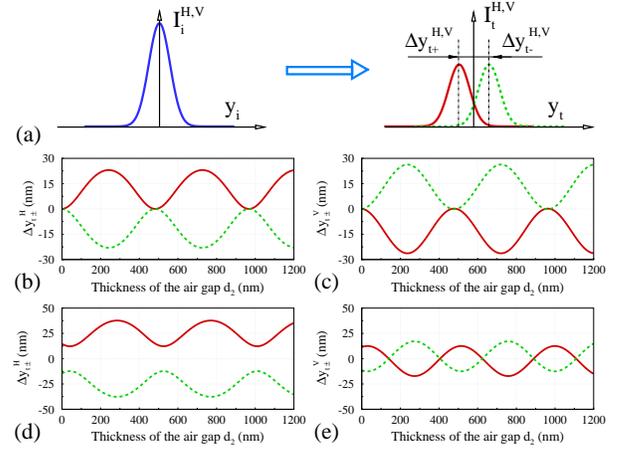}
\caption{\label{Fig3} (Color online) (a) Schematic illustrating the
spin-dependent transverse splitting. The transverse displacements
versus the thickness of the air gap $d_2$ in the situation of
transmission with incident angle $\theta_i=\pi/6$. (b) and (c) show
the transverse displacements of $H$ and $V$ components,
respectively, for the symmetric system. (d) and (e) also represent
the transverse displacements but for the asymmetric system.}
\end{figure}

The spin-dependent splitting in SHE of light is schematically shown
in Fig.~\ref{Fig3}(a). From Eqs.~(\ref{TSRH}) and~(\ref{TSRV}), we
know that the transverse displacements of $H$ and $V$ polarizations
would have just opposite tendency versus $d_2$ which can also be
seen from Fig.~\ref{Fig3}(b)-\ref{Fig3}(e). For a fixed incident
angle of $\theta_i=\pi/6$, the transverse displacements present
sine-like oscillations since the value of $|t_s|/|t_p|$ or
$|t_p|/|t_s|$ is periodic due to the Fabry-Perot resonance in the
layered structures. In the symmetric case, the transverse
displacements present a sine-like oscillation in the range of
zero-positive-zero or zero-negative-zero for a certain polarization
component [Fig.~\ref{Fig3}(b)-\ref{Fig3}(c)]. In the asymmetric
case, for $H$-polarization, the transverse displacements can exhibit
a positive and negative transverse displacement for left- and
right-polarized components, respectively [Fig.~\ref{Fig3}(d)]. It is
interesting to note that the transverse displacements exhibits a
sine-like oscillation in the range of negative-zero-positive values
for $V$-polarization, which can be modulated via tuning the
thickness of the air gap [Fig.~\ref{Fig3}(e)]. It indicates that the
SHE of light can be greatly enhanced or suppressed, or even
completely eliminated.

From the above analysis, we know that the transverse displacements
are related to the ratio between the Fresnel transmission
coefficients $|t_p|$ and $|t_s|$, whose dependence on the thickness
of the air gap are periodic due to the Fabry-Perot resonance in the
layered nanostructures. Hence, we can expect that a nanostructure
with a large ratio of $|t_s|/|t_p|$ or $|t_p|/|t_s|$ would extremely
enhance the SHE of light. On the contrary, a small ratio of
$|t_s|/|t_p|$ or $|t_p|/|t_s|$ would greatly suppress the SHE of
light. In fact, the phase difference $\phi_p-\phi_s$ of the Fresnel
transmission coefficients may also be used to modulate the SHE of
light since it can change the sign of $\Delta y_{t\pm}$. The
nanostructures with a metamaterial layer whose refractive index can
be tailored arbitrarily can be a good candidate to support this
prediction~\cite{Luo2009}. Under the condition of
$|t_s|/|t_p|\cos(\phi_s-\phi_p)=\eta$ ($H$ polarization) and
$|t_p|/|t_s|\cos(\phi_p-\phi_s)=\eta$ ($V$ polarization), the SHE of
light can be suppressed completely.

It should be mentioned that the spatial separation of the spin
components can also be tuned continuously by varying the incident
angle in a single air-glass interface~\cite{Hosten2008}. However,
the refracted angle changes and the transmission coefficients
decrease accordingly as the incident angle increases. Hence, it is
disadvantage for potential application to nano-photonic devices. As
shown in above, the wave packet in our scheme is incident at a fixed
angle, and the transverse displacements can be tuned to either a
negative or a positive value, or even zero, by just adjusting the
structure parameters. Meanwhile, the wave packet in the layered
nanostructure exhibit much higher transmission coefficients than in
the single air-glass interface. Hence, the layered nanostructures
provide more flexibility for modulating the SHE of light. We will
search for suitable nanostructures to manipulate it in the future.
It is expected that the SHE of light in layered nanostructures will
be useful for designing very fast optical switches, for example, by
replacing the air gap by material whose refractive index can be
tuned by a electric field.

\section{Conclusions}
In conclusion, we have revealed a tunable SHE of light in layered
nanostructures. From the viewpoint of classical electrodynamics, we
have established a general propagation model to describe the
spin-dependent transverse splitting of wave packet in the SHE of
light. By modulating the structure parameters, the transverse
displacements exhibit tunable values ranging from negative to
positive, including zero, which means that the SHE of light can be
greatly enhanced or suppressed, or completely eliminated. We have
shown that the physical mechanism underlying this intriguing
phenomenon is the optical Fabry-Perot resonance in the layered
nanostructure. These findings provide a pathway for modulating the
SHE of light, and thereby open the possibility for developing new
nano-photonic devices.

\begin{acknowledgements}
This research was partially supported by the National Natural
Science Foundation of China (61025024, 11074068, and 10904036).
\end{acknowledgements}

\appendix
\section{Calculation of the reflected and transmitted angular spectra} \label{AppA}
In this appendix we give a detailed calculation of the reflected and
transmitted angular spectra. From the central frame $x_iy_iz_i$ to
the local frame $X_iY_iZ_i$, the following three steps should be
carried out. First, we transform the electric field from the
reference frame $x_iy_iz_i$ around the $y$ axis by the incident
angle $\theta_i$ to the frame $xyz$:
$\tilde{E}_{xyz}=m_{{x_iy_iz_i}\rightarrow{xyz}}\tilde{E}_{x_iy_iz_i}$,
where
\begin{eqnarray}
m_{{x_iy_iz_i}\rightarrow{xyz}}=\left[
\begin{array}{ccc}
\cos\theta_i & 0 & -\sin\theta_i\\
0 & 1 & 0\\
\sin\theta_i & 0 & \cos\theta_i
\end{array}
\right].\label{matrixr}
\end{eqnarray}
Then, we transform the electric field from the reference frame $xyz$
around the $y$ axis by an angle $k_{iy}/(k_0\sin\theta_i)$ to the
frame $XYZ$, and the correspondingly matrix is given by
\begin{eqnarray}
m_{{xyz}\rightarrow{XYZ}}=\left[
\begin{array}{ccc}
1 & \frac{k_{iy} }{k_0\sin\theta_i} & 0\\
-\frac{k_{iy}}{k_0\sin\theta_i} & 1 & 0\\
0 & 0 & 1
\end{array}
\right],\label{matrixr}
\end{eqnarray}
where $k_0$ is the wave number in vacuum. Finally, we transform the
electric field from the reference frame $XYZ$ around the $y$ axis by
an angle $-\theta_i$ to the frame $X_iY_iZ_i$, and the matrix can be
written as
\begin{eqnarray}
m_{{XYZ}\rightarrow{X_iY_iZ_i}}=\left[
\begin{array}{ccc}
\cos\theta_i & 0 & \sin\theta_i\\
0 & 1 & 0\\
-\sin\theta_i & 0 & \cos\theta_i
\end{array}
\right].\label{matrixr}
\end{eqnarray}
Thus, the rotation matrix from the central frame $x_iy_iz_i$ to the
local frame $X_iY_iZ_i$ can be written as
$M_{{x_iy_iz_i}\rightarrow{X_iY_iZ_i}}=m_{{XYZ}\rightarrow{X_iY_iZ_i}}
m_{{xyz}\rightarrow{XYZ}}m_{{x_iy_iz_i}\rightarrow{xyz}}$, and we
have
\begin{eqnarray}
M_{{x_iy_iz_i}\rightarrow{X_iY_iZ_i}}=\left[
\begin{array}{ccc}
1 &\frac{k_{iy} \cot\theta_i}{k_0} &0 \\
-\frac{k_{iy} \cot\theta_i}{k_0} & 1 & \frac{k_{iy}}{k_0} \\
0 & -\frac{k_{iy}}{k_0} & 1
\end{array}
\right].\label{matrixr}
\end{eqnarray}
For an arbitrary wave vector, the reflected field is determined by
$\tilde{E}_{X_rY_rZ_r}=r_{p,s}\tilde{E}_{X_iY_iZ_i}$, where $r_p$
and $r_s$ are the Fresnel reflection coefficients. The reflected
field should be transformed from $X_rY_rZ_r$ to $x_ry_rz_r$.
Following the similar procedure, the reflected field can be obtained
by carrying out three steps of transformation:
$\tilde{E}_{x_ry_rz_r}=M_{{X_rY_rZ_r}\rightarrow{x_ry_rz_r}}\tilde{E}_{X_rY_rZ_r}$
where
\begin{eqnarray}
M_{{X_rY_rZ_r}\rightarrow{x_ry_rz_r}}=\left[
\begin{array}{cc}
1 &\frac{k_{ry} \cot\theta_i}{k_0} \\
-\frac{k_{ry} \cot\theta_i}{k_0} & 1
\end{array}
\right].\label{matrixr}
\end{eqnarray}
Here, only the two-dimensional rotation matrices is taken into
account, since the longitudinal component of electric field can be
obtained from the divergence equation
$\tilde{E}_{az}k_{az}=-(\tilde{E}_{ax}k_{ax}+\tilde{E}_{ay}k_{ay})$.
The reflection matrix can be written as
\begin{eqnarray}
M_R=M_{{X_rY_rZ_r}\rightarrow{x_ry_rz_r}}\left[
\begin{array}{cc}
r_p &0 \\
0 & r_s
\end{array}
\right]M_{{x_iy_iz_i}\rightarrow{X_iY_iZ_i}}.\label{matrixrI}
\end{eqnarray}
The reflected angular spectrum is related to the boundary
distribution of the electric field by means of the relation
$\tilde{E}_r(k_{rx},k_{ry})=M_R\tilde{E}_i(k_{ix},k_{iy})$, and we
have
\begin{eqnarray}
\left[\begin{array}{cc}
\tilde{E}_r^H\\
\tilde{E}_r^V
\end{array}\right]
=\left[
\begin{array}{cc}
r_p&\frac{k_{ry} (r_p+r_s) \cot\theta_i}{k_0} \\
-\frac{k_{ry} (r_p+r_s)\cot\theta_i}{k_0} & r_s
\end{array}
\right]\left[\begin{array}{cc}
\tilde{E}_i^H\\
\tilde{E}_i^V
\end{array}\right],\label{matrixrII}
\end{eqnarray}

We proceed to consider the transmitted field. Following the similar
procedure, we obtain the transform matrix from $X_tY_tZ_t$ to
$x_ty_tz_t$ as
\begin{eqnarray}
M_{{X_tY_tZ_t}\rightarrow{x_ty_tz_t}}=\left[
\begin{array}{cc}
1 &-\frac{k_{ty} \cos\theta_t}{k_0 \sin\theta_i} \\
\frac{k_{ty} \cos\theta_t}{k_0 \sin\theta_i} & 1
\end{array}
\right],\label{matrixr}
\end{eqnarray}
where $\theta_t$ is the transmitted angle. For an arbitrary wave
vector, the transmitted field is determined by
$\tilde{E}_{X_tY_tZ_t}=t_{p,s}\tilde{E}_{X_iY_iZ_i}$, where $t_p$
and $t_s$ are the Fresnel transmission coefficients. Hence, the
transmission matrix can be written as
\begin{eqnarray}
M_T=M_{{X_tY_tZ_t}\rightarrow{x_ty_tz_t}}\left[
\begin{array}{cc}
t_p &0 \\
0 & t_s
\end{array}
\right]M_{{x_iy_iz_i}\rightarrow{X_iY_iZ_i}}.\label{matrixtI}
\end{eqnarray}
The transmitted angular spectrum is related to the boundary
distribution of the electric field by means of the relation
$\tilde{E}_t(k_{tx},k_{ty})=M_T\tilde{E}_i(k_{ix},k_{iy})$, and can
be written as
\begin{eqnarray}
\left[\begin{array}{cc}
\tilde{E}_t^H\\
\tilde{E}_t^V
\end{array}\right]
=\left[
\begin{array}{cc}
t_p&\frac{k_{ty} (t_p-\eta t_s) \cot\theta_i}{k_0} \\
\frac{k_{ty} (\eta t_p-t_s)\cot\theta_i}{k_0} & t_s
\end{array}
\right]\left[\begin{array}{cc}
\tilde{E}_i^H\\
\tilde{E}_i^V
\end{array}\right],\label{matrixtII}
\end{eqnarray}
where $\eta=\cos\theta_t/\cos\theta_i$.

\section{Transverse displacements for arbitrary linear polarization}\label{AppB}
For an arbitrary linearly polarized beam, the transverse
displacements of the reflected field are given by
\begin{equation}
\Delta y_{r\pm}=\cos^2\gamma_r\Delta y_{r\pm}^H+\sin^2\gamma_r\Delta
y_{r\pm}^V\label{TSRHV},
\end{equation}
where $\gamma_r$ is the reflected polarization angle. In the frame
of classical electrodynamics, the reflection polarization angle is
determined by:
\begin{equation}
\cos\gamma_r=\frac{\cos\gamma_i
\mathrm{Re}[r_p]}{\sqrt{\cos^2\gamma_i\mathrm{Re}[r_p]^2+\sin^2\gamma_i\mathrm{Re}[r_s]^2}},
\end{equation}
\begin{equation}
\sin\gamma_r=\frac{\sin\gamma_i
\mathrm{Re}[r_s]}{\sqrt{\cos^2\gamma_i\mathrm{Re}[r_p]^2+\sin^2\gamma_i\mathrm{Re}[r_s]^2}}.
\end{equation}
Here, $\gamma_i$ is the incident polarization angle. For an
arbitrary linearly polarized wave-packet, the transverse
displacements of the transmitted field are given by
\begin{equation}
\Delta y_{t\pm}=\cos^2\gamma_t\Delta y_{t\pm}^H+\sin^2\gamma_t\Delta
y_{t\pm}^V\label{TSTHV},
\end{equation}
where the transmission polarization angle $\gamma_t$ determined by
\begin{equation}
\cos\gamma_t=\frac{\cos\gamma_i
\mathrm{Re}[t_p]}{\sqrt{\cos^2\gamma_i\mathrm{Re}[t_p]^2+\sin^2\gamma_i\mathrm{Re}[t_s]^2}}\label{COST},
\end{equation}
\begin{equation}
\sin\gamma_t=\frac{\sin\gamma_i
\mathrm{Re}[t_s]}{\sqrt{\cos^2\gamma_i\mathrm{Re}[t_p]^2+\sin^2\gamma_i\mathrm{Re}[t_s]^2}}\label{SINT}.
\end{equation}

\end{document}